\documentclass[journal=jctcce,manuscript=article,layout=twocolumn]{achemso}

\usepackage{graphicx}
\usepackage{subcaption}
\usepackage{pgfplots}
\pgfplotsset{compat=1.18}
\usepackage{xcolor}
\usepackage{multirow}
\usepackage[hidelinks]{hyperref}
\usepackage{braket}
\usepackage{amsmath}
\usepackage{amssymb}
\usepackage{bbold}
\usepackage{booktabs}
\usepackage[exponent-product = \cdot]{siunitx}

\usepackage{pdfpages}
\numberwithin{equation}{section}
\usepackage[version=3]{mhchem}
\usepackage{tikz}
\usepackage{adjustbox}
\usepackage{appendix}
\usepackage{bm}
\usepackage{dsfont}
\usepackage{nicematrix}
\usepackage{notes2bib}

\DeclareUnicodeCharacter{2212}{-}

\author{Mike Keizer}
\affiliation{Department of Chemistry and Pharmaceutical Sciences, Vrije Universiteit, De Boelelaan 1105, 1081 HV Amsterdam, The Netherlands}

\author{Stefano Paggi}
\affiliation{
Univ. Toulouse, CNRS, LPT, and European Theoretical Spectroscopy Facility (ETSF), Toulouse, France}
\author{J. Arjan Berger}
\affiliation{Univ. Toulouse, CNRS, LCPQ, and European Theoretical Spectroscopy Facility (ETSF), Toulouse, France}
\author{Pina Romaniello}
\affiliation{
Univ. Toulouse, CNRS, LPT, and European Theoretical Spectroscopy Facility (ETSF), Toulouse, France}

\author{Arno F\"orster}
\email{a.t.l.foerster@vu.nl}
\affiliation{Department of Chemistry and Pharmaceutical Sciences, Vrije Universiteit, De Boelelaan 1105, 1081 HV Amsterdam, The Netherlands}

\title{Benchmark of Multi-Channel Dyson Equation and Algebraic Diagrammatic Construction Methods for molecules}

\begin{document}

\begin{abstract}
The Dyson-algebraic diagrammatic construction (ADC) and the multi-channel Dyson equation (MCDE) formalisms explicitly leverage multi-particle channels to formulate correlated theories of the single-particle Green's function that produce positive semi-definite spectral functions by construction. 
% In this work, we make explicit connections between both formalisms and show that the MCDE formalism is completely equivalent to the extended ADC(2) [ADC(2)-X] approximation. 
While the MCDE is strictly rooted in the Dyson formalism, most ADC calculations are performed in the non-Dyson (nD) framework that decouples electron attachment and detachment sectors. We benchmark the Dyson-ADC(2)-X [that is equivalent to the (3,1)-MCDE] and ADC(3) approximations on a set of 58 ionization potentials of 23 small molecules for which near-full configuration interaction reference data exist. Comparison of Dyson- to nD-ADC(3) reveals deviations of the order of 0.1 eV between both methods, calling into question the reliability of the nD approximation. We show that Dyson-ADC gives similar accuracy for first IPs as for semi-valence and semi-core transitions. Finally, we also benchmark the screened (3,1)-MCDE that screens all ladder interactions, and show that it improves over its unscreened counterpart.
\end{abstract}

\section{Introduction}

Higher-order Green's functions are important conceptual tools to formulate correlated theories of the single-particle Green's function $G_1$.\cite{Stefanucci2013NonequilibriumIntroduction, martin2016} The Martin-Schwinger hierarchy\cite{Martin1959} links it to the 2-particle Green's function $G_2$, which in turn is expressed through the three-particle Green's function $G_3$, and so on. The formulation of approximations to $G_1$ often starts with the definition of the irreducible self-energy $\Sigma$, and the reformulation of the EOM for $G_1$ as a Dyson equation. 

Very often, for instance, in Hedin's equations,\cite{Hedin1965, Aryasetiawan1998} or in Parquet theory,\cite{DeDominicis1964, Bickers1989, Bickers2004, Marie2025ParquetApproximation} only 1- and 2-body quantities are used explicitly to calculate $\Sigma$. However, the 3-particle Green's function is another important conceptual tool to derive expressions for the single-particle self-energy. As shown by \citet{Ethofer1969TheProblem} and \citet{Winter1972StudyFunction}, the single-particle self-energy can be expressed in terms of the 3-particle response function (obtained from $G_3$ by subtracting certain disconnected terms), which in turn obeys an integral equation containing an effective 3-particle vertex. For instance, the second-order Green's function (GF2) self-energy\cite{Abrikosov1975, Szabo2012} is obtained by approximating the 3-particle response by its zeroth-order contribution. Physically, this self-energy describes the coupling of a particle to non-interacting 2-particle-1-hole (2p1h) and 2-hole-1-particle (2h1p) propagation channels. Also the $GW$- or $T$-matrix approximations can be upfolded to the 2p1h and 2h1p channels.\cite{Bintrim2021, Monino2022, Monino2023, Tolle2023} We, therefore, conclude that these approximations can also be directly linked to the 3-particle Green's function.

Also the (Dyson) algebraic diagrammatic construction (ADC) method,\cite{Schirmer1983, vonNiessen1984ComputationalFunction} building on the diagrammatic self-energy resummation techniques of Cederbaum\cite{Cederbaum1973, Cederbaum1975, Cederbaum1977}, expresses the self-energy in terms of the 3-particle response and uses perturbation theory in the Coulomb interaction to arrive at a hierarchy of systematically improvable approximations that obey the spectral representation of the exact self-energy.\cite{Winter1972StudyFunction, vonNiessen1984ComputationalFunction} To lowest order in the Coulomb interaction, the ADC(2) approximation is obtained, which is strictly equivalent to GF2, while at third order, the ADC(3) approximation is obtained, which is equivalent to the extended 2ph-Tamm-Dancoff approximation (TDA).\cite{walter1981two} The intermediate ADC(2)-X approximation is equivalent to the 2ph-TDA.\cite{Cederbaum1977,cederbaum1977theoretical,schirmer1978two} In both methods, TDA refers to the decoupling of the 2p1h and 2h1p sectors of the three-particle Green's function, a defining feature of ADC theory (for a formulation with these sectors coupled see Ref.~\citenum{Degroote2011}). However, in modern formulations, this connection to the three-particle Green's function is typically left implicit. Instead, ADC is typically presented as a self-contained perturbative construction of the self-energy or, equivalently, through the intermediate state representation (ISR)\cite{Schirmer1991Closed-formMatrices} in which the three-particle character of the underlying propagator is implicit in the choice of excitation manifold rather than appearing as an explicit Dyson equation. 

The ADC self-energies are guaranteed to admit a Lehmann representation, which in turn guarantees that the resulting spectral function is positive semi-definite (PSD), an important property that many diagrammatic methods do not fulfil.\cite{Cederbaum1975, Stefanucci2014, Bruneval2025GW+2SOSEXSemidefinite} A direct consequence is that the solution of the ADC equations up to third order\bibnote{Starting from fourth-order, the configuration space must be enlarged, and the self-energy also includes couplings to the 3-particle-2-hole (3p2h) and 3-hole-2-particle (3h2p) sectors. See Ref.~\citenum{Schirmer1983}.} can be recast as the solution of a Hermitian eigenproblem through upfolding to the 2p1h and 2h1p sectors from the single-particle space.\cite{Bruneval2025GW+2SOSEXSemidefinite} 
This property, in principle, allows for obtaining spectral properties through a single diagonalization. On the other hand, diagonalization of the resulting large matrix quickly becomes intractable. Iterative solvers based on the standard Davidson method\cite{Davidson1975} can efficiently extract low-lying eigenvalues of these matrices. In most applications, where one is interested in a few principal ionization potentials (IP) or electron affinities (EA), this is rather inefficient since they reside in the middle of the spectrum of the ADC Hamiltonian, and specialized solvers are required.\cite{Weikert1996BlockFunction}

This problem can be overcome by non-Dyson (nD)-ADC methods that decouple the IP and EA sectors of the ADC Hamiltonian by applying the ADC procedure to the greater and lesser self-energies separately.\cite{Schirmer1998APropagator} Since the desired eigenvalues now reside at the margins of the IP- and EA-Hamiltonians, standard Davidson solvers can be used. For this reason, and also since nD- and Dyson ADC are believed to give comparable results,\cite{Trofimov2005MolecularApproach} nD-ADC is usually preferred in practice. It has become a widespread method to calculate quasiparticle (QP) energies in molecules\cite{
Schneider2015AStudy,
Dempwolff2019EfficientMolecules, 
Dempwolff2020IntermediateImplementation,
Herbst2020Adcc:Methods,
Leitner2024Fourth-OrderMethods,
Banerjee2019Third-orderImplementation,
Banerjee2023AlgebraicSpectra,
Majumder2025AlgebraicCorrelation}
and has also been implemented for solids.\cite{Banerjee2022Non-DysonSolids} 

Unlike Parquet theory or Hedin's equations, which are in principle formulated through the self-consistent update of the single-particle Green's function through the Dyson equation, the standard formulation of ADC assumes a Hartree--Fock (HF) Green's function throughout. Therefore, ADC typically calculates the static part of the self-energy, $\Sigma(\infty)$, which captures the instantaneous interactions of an electron with the surrounding many-body system, with a correlated density matrix.  $\Sigma(\infty)$ can be expanded order-by-order in the Coulomb interaction,\cite{Schirmer1998APropagator} yet more involved schemes that take into account higher-order terms and orbital relaxation have been developed\cite{vonNiessen1984ComputationalFunction, Schirmer1989OnValues, Trofimov2005MolecularApproach} and are often employed in practice.\cite{Dempwolff2020IntermediateImplementation} The update of the Dyson equation in (QP)\cite{Faleev2004, VanSchilfgaarde2006, Kotani2007} self-consistent theories takes care of the static self-energy part automatically. However, in the non-self-consistent approximations\cite{Hybertsen1985, Hybertsen1986, Golze2019, Marie2024b} that are commonly employed to perform (vertex-corrected) $GW$ calculations, density updates must be reintroduced.\cite{Bruneval2019a} This is usually not done, since tuning of the starting point can compensate for this and often leads to excellent results for first IPs.\cite{Knight2016, Caruso2016, Bruneval2021a, McKeon2022} However, when HF orbitals are used, the update of the static self-energy contribution, for instance through the so-called $GW$-linearized density matrix,\cite{Bruneval2019, Bruneval2019a, Bruneval2021} is absolutely crucial for accurate results.\cite{Bruneval2021a, Tolle2026ConnectionCluster,Pavlyukh2026ApproachingSelf-Energies}

More recently, the (3,1)-multi-channel Dyson equation [(3,1)-MCDE] method has been formulated as a theory that utilizes the three-body Green's function to express the single-particle self-energy.\cite{Riva2022, Riva2023, Riva2024, Riva2025, Paggi2025GroundEquation} In contrast to ADC, the (3,1)-MCDE approach hence starts directly from the Dyson equation for $G_3$, so that the propagation of an added or removed electron together with the particle-hole pair it excites is treated as a single, three-particle propagation problem. Despite these different constructions, both formalisms converge to the same working equations in practice.\cite{Demartini2026AlgebraicEquation} However, MCDE calculations reported so far have not updated the static self-energy and instead left it constant at the HF level. The MCDE formalism opens up new avenues for incorporating screening of the electron-electron interaction into the theory that recently led to the development of the screened (3,1)-MCDE (sMCDE).\cite{Romaniello2026DirectEquation} 

The (3,1)-MCDE has recently been used to calculate photoemission spectra of atoms and small molecules.\cite{Paggi2026CoreEquation} However, a direct benchmark of MCDE and ADC methods for molecules is missing so far. To this end, we here express the MCDE and ADC formalisms in a unified Green's function-based representation that highlights their connections. We then compare Dyson- and nD-ADC(2)(-X), sMCDE, and ADC(3) approximations for a set of 58 IPs of 23 small molecules for which near full configuration interaction (FCI) reference values exist,\cite{Marie2024} paying special attention to the density matrix with which the static self-energy contribution is evaluated. We also introduce a root-following harmonic Davidson solver\cite{Huang2022TwoPair} combined with an efficient linearization procedure that allows for the efficient calculation of selected IPs in Dyson-ADC methods.

\section{Theory}
\subsection{The three-body Green's function}
For the following discussion to be self-contained, we here present the ADC and the (3,1)-MCDE working equations from a unified perspective. At zero temperature, the propagation of three fermions in an $N$-electron system is described by the time-ordered three-body Green's function (3-GF), defined by\cite{Riva2022}
\begin{equation}
\begin{split}
G&_3(1,2,3,1',2',3')
\\
&= -i \Big\langle \Psi_0^{\text{N}} \Big| 
\mathcal{T}\Big[
\psi(1)\psi(2)\psi(3) \\
& \ \ \ \ \ \times\psi^\dagger(1')\psi^\dagger(2')\psi^\dagger(3') \Big]
\Big|\Psi_0^{\text{N}}\Big\rangle,
\end{split}
\end{equation}
where $|\Psi_0^{\text{N}}\rangle$ is the N-electron ground state of the system, and $\psi^{\dagger}$ and $\psi$  are the creation and annihilation Heisenberg operators, respectively. The notation $(1) = (x_1,t_1) = (r_1, s_1, t_1)$ refers to the space, spin and time coordinates. Wick's time-ordering operator generates all possible orderings of the field operators, corresponding to different propagation channels. However, for photoelectron spectroscopy processes only the 2p1h and 2h1p sectors are relevant, describing the propagation of an added or removed electron accompanied by an electron-hole pair excitation.\cite{Riva2023, Riva2024} Neglecting the time differences between operators within each three-particle cluster then yields the 2p1h/2h1p three-body Green's function\cite{Ethofer1969TheProblem, Riva2024}
\begin{equation}
\begin{split}
G^{\text{I/II}}_3&(\{x\},\{x'\}; t-t') \\
&= -i \Big\langle \Psi_0^{\mathrm N} \Big|
\mathcal{T}\Big\{
\left[\psi^\dagger(x_{3'})\psi(x_2)\psi(x_1)\right]_t
\\
&\ \ \ \times
\left[\psi^\dagger(x_{1'})\psi^\dagger(x_{2'})\psi(x_3)\right]_{t'}
\Big\}
\Big|\Psi_0^{\mathrm N}\Big\rangle \;,
\end{split}
\end{equation}
that describes the simultaneous propagation of three particles. For convenience we adopted the notation $\{x\} = \{x_1, x_2, x_3\}$, and the group of operators $\left[\psi^\dagger(x_{3'})\psi(x_2)\psi(x_1)\right]_t$, known as intermediate state operators,\cite{Schirmer1991Closed-formMatrices, Dreuw2023AlgebraicInterpretation} acts at the same time $t$.  It will be convenient to project $G_3$ onto a one-electron spin orbital basis set $\{\phi_i\}$ that diagonalizes the non-interacting 3-GF $G_3^0$. By additionally performing a Fourier transform in $\tau = t - t'$, the 3-GF becomes\cite{Riva2024}
\begin{equation}
\begin{split}
    G^{\text{\text{I/II}}}_{ijl;mok}(\omega) = &\int d\{x_q\}d\{x_{q'}\}\phi_i^*(x_1)\phi_j^*(x_2) \\
    & \times \phi_l(x_{3'})\phi_m(x_{1'})\phi_o(x_{2'})\phi_k^*(x_3) \\
    & \times G_3^{\text{\text{I/II}}}(\{x_q\}, \{x_{q'}\}; \omega),
\end{split}
\end{equation}
where the superscripts I and II conveniently abbreviate the 2p1h and 2h1p spaces, respectively. The Lehmann representation is obtained by inserting a complete basis set of Fock eigenstates, $\mathds{1} = \sum_n |\Psi_n^{\text{N}\pm 1}\rangle \langle \Psi_n^{\text{N} \pm 1} |$, writing the 3-GF as\cite{Riva2022}
\begin{equation}
\begin{split}
    G^{\text{\text{I/II}}}_{ijl;mok}(\omega) &= \sum_n \frac{X_n^{ijl }\widetilde{X}_n^{mok}}{\omega - (E_n^{\text{N+1}} - E^{\text{N}}_0) + i\eta} \\
    &+ \sum_n \frac{Z_n^{ijl}\widetilde{Z}_n^{mok}}{\omega - (E^{\text{N}}_0 - E_n^{\text{\text{N-1}}}) - i\eta},
\end{split}
\end{equation}
with
\begin{equation}
\begin{split}
    X_n^{ijl} &=  \big\langle \Psi_{0}^{\text{N}} \big|\hat{c}^{\dagger}_l\hat{c}_j\hat{c}_i\big |\Psi_n^{\text{N+1}}\big\rangle, \\ \widetilde{X}_n^{mok} &= \big\langle \Psi_n^{\text{N+1}}\big|\hat{c}^{\dagger}_m\hat{c}^{\dagger}_o\hat{c}_k \big{|}\Psi_{0}^{\text{N}}\big\rangle,  \\
    Z_n^{ijl} &= \big\langle \Psi_n^{\text{\text{N-1}}} \big| \hat{c}^{\dagger}_l \hat{c}_j \hat{c}_i  \big| \Psi_0^{\text{N}} \big \rangle, \\
    \widetilde{Z}_n^{mok} &=  \big\langle \Psi_0^{\text{N}} \big| \hat{c}^{\dagger}_m \hat{c}^{\dagger}_o \hat{c}_k \big| \Psi_n^{\text{\text{N-1}}}\big\rangle,
\end{split}
\end{equation}
where $\hat{c}^{\dagger}$ and $\hat{c}$ are respectively the one-electron spin orbital creation and annihilation operators. The spectral representation thus reveals the correspondence between the pole structure of the 3-GF and the electron addition and removal energies in the N-body system. Moreover, the pole residues are the products of transition amplitudes, and relate to the spectral intensities of the electron-attachment and ionization experiments.\cite{cederbaum1977theoretical, Riva2022} Even though the Lehmann representation provides an explicit relation to excitation energies, its direct evaluation is impractical since it requires the exact many-body eigenstates. 

\subsection{The one-body Green's function}

The interacting one-body Green's function is obtained from the reference Green's function $G^0$ through the Dyson equation, which reads\cite{Abrikosov1975}
\begin{equation}
    G(\omega) = G^0(\omega) + G^0 (\omega) \Sigma_c(\omega)G(\omega) \;.
    \label{2.6}
\end{equation}
Here and in the following, we assume that $G^0(\omega)$ is the Hartree--Fock (HF) Green's function. The self-energy $\Sigma$ contains all many-body correlated interactions and can be further decomposed into\cite{Schirmer1983}
\begin{equation}
\label{eq:sigma_static}
    \Sigma_c(\omega) = \Sigma_c(\infty) + M(\omega),
\end{equation}
with $\Sigma_c(\infty)$ and $M(\omega)$ the static and dynamical part, respectively. 
For a fixed HF reference, the static self-energy can be represented in terms of the one-particle Green's function, according to\cite{cederbaum1977theoretical}
\begin{equation}
\label{eq:sigma_infty}
    \begin{split}
        \Sigma_{c_{pq}}(\infty) &=\frac{1}{2\pi i}\sum_{rs} \langle pr||qs \rangle \\
        & \qquad \times \oint e^{i\omega 0^+} \left[G_{rs}(\omega) - G^{0}_{rs}(\omega)\right] d\omega,
    \end{split}
\end{equation}
where the contour integral closes over the upper half-plane. $G^0$ must be subtracted to avoid double counting the HF density matrix. Here, $\langle pq||rs \rangle = \langle pq|rs \rangle - \langle pq|sr \rangle $ are the anti-symmetrized Coulomb integrals in physicist notation, with
\begin{equation}
    \langle pq|rs \rangle = \int dx_1 dx_2 \frac{\phi^*_p(x_1) \phi^*_q(x_2) \phi_r(x_1) \phi_s(x_2)}{|r_1 - r_2|}
\end{equation}

To determine the dynamical part of the self-energy, we first rewrite Eq.~\eqref{2.6} as
\begin{equation}
    G(\omega) = \left[(G^{0}(\omega))^{-1} - \Sigma_c(\omega)\right]^{-1}.
\end{equation}
To solve this equation with standard numerical techniques, we introduce an auxiliary many-body Hilbert space where we solve the Dyson equation through the equivalent Hermitian and frequency-independent eigenvalue problem\cite{Schirmer1991Closed-formMatrices, Riva2024}
\begin{equation}
    H_{\text{eff}} X = E X \;,
\end{equation}
with the effective many-body Hamiltonian given by
\begin{equation}
    H^{\text{D}}_{\text{eff}} = 
        \begin{pNiceArray}{c|cccc}
        \\
        \epsilon + \Sigma_c(\infty) & & U^{\text{I}} & & U^{\text{II}} \\ & \\
        \hline \\ \left(U^{\text{I}}\right)^{\dagger} & & {K^{\text{I}}} + {C^\text{I}} & & 0 \\  \\  \left(U^{\text{II}}\right)^{\dagger}
         & & 0  & & {K^{\text{II}}} + {C^{\text{II}}} \\ \\
        \end{pNiceArray},
        \label{2,1:effectiveHamiltonian}
\end{equation}
where the superscript D represents the Dyson formalism. The effective Hamiltonian consists of a one-body sector coupled to the many-body environment. Focusing on the 3-body Green's function, the one-body block contains the QP energies, while the 3-body sector contains the satellites and represents the 2p1h/2h1p configuration space responsible for the dynamical 1-body self-energy. 3p2h/3h2p and higher configurations can additionally be folded into the 3-particle configuration space. The diagonal matrices $\epsilon$ and $K$ consist of the non-interacting one-body and three-body energies, respectively, that in turn represent non-interacting QPs and satellites. The Hermitian interaction matrix $C$ describes the three-particle interactions, and the vertex $U$ couples the one-body block to the three-body reservoir. Different many-body Green's function methods correspond to different approximations to the effective Hamiltonian, derived from distinct perturbative contributions to the dynamical self-energy, and consequently, to the one- and three-body blocks.\cite{Schirmer1983, Riva2023}
\subsection{ADC \& MCDE}
Although the (3,1)-MCDE and ADC formalisms employ the same configuration space, they differ fundamentally in their initial construction. In (3,1)-MCDE, the multi-channel Dyson equation is introduced in the direct sum space $\mathcal{H}_{\text{1p}} \bigoplus \mathcal{H}_{\text{3p}}$ where it connects the interacting and non-interacting multichannel 3-GF through the multi-channel self-energy (MCSE)\cite{Riva2023, Riva2024}
\begin{equation}
    \mathcal{G}_3(\omega) = \mathcal{G}^0_3(\omega) + \mathcal{G}_3^0(\omega)\Sigma_3(\omega)\mathcal{G}_3(\omega) \;.
\end{equation}
Here, after removal of redundant information,\cite{Riva2024}
\begin{equation}
\mathcal{G}^0_3(\omega)=
\begin{pNiceArray}{cc}
G^0_1(\omega) & 0\\
0 & G^{0,3p}(\omega)
\end{pNiceArray} \;.
\end{equation}
A numerical evaluation of the MCDE requires an approximation of $\Sigma_3$. This is achieved by restricting the MCSE to static, pairwise direct and exchange interactions, corresponding to the RPAx approximation, and writing $\Sigma_3$ as\cite{Riva2022, Riva2023, Riva2024, Riva2025}
\begin{equation}
\Sigma_3 = 
    \begin{pNiceArray}{cc}
    0 & \Sigma^c \\
    (\Sigma^c)^{\dagger} & \Sigma^{3p}
    \end{pNiceArray}.
\end{equation}
Since $\mathcal{G}_3^0$ is exact up to first order, the static mean-field contribution vanishes ($\Sigma^{1p} = \Sigma_c(\infty) = 0$) to this order, and all correlation effects are captured through the coupling block $\Sigma^c$ and the $\Sigma^{3p}$ block. The missing contributions to $\Sigma_c(\infty)$ beyond first order can either be introduced explicitly, by using a correlated one-body GF (see Section~\ref{sec:static_theory}), or through self-consistency.

The explicit expressions for the MCSE blocks are found by analyzing the electron-hole and particle-particle channels of the Bethe-Salpeter equation (BSE) kernel.\cite{Riva2023,Riva2024,Riva2025, Marie2025AnomalousEquation}
\\
In contrast to the (3,1)-MCDE formalism, ADC starts from the one-body Dyson equation and formulates the dynamical self-energy in terms of the three-body response function, defined as\cite{Ethofer1969TheProblem, Winter1972StudyFunction, Schirmer1983}
\begin{equation}
\begin{aligned}
    & \Pi_{ijl;mok}^{\text{\text{I/II}}}(\omega) = \\
    & \left[G_{ijl;mok}^{\text{\text{I/II}}} - \sum_{pq}G_{jl;kp}G^{-1}_{pq}G_{qi;mo}\right](\omega),
\end{aligned}
\end{equation}
such that the (1-body) dynamical self-energy reads 
\begin{equation}
    M_{pq}(\omega) = \sum_{i<jl;\,m<ok}\langle pl||ij \rangle^{*}\,
    \Pi^{\text{\text{I/II}}}_{ijl;mok}(\omega)\,\langle qk||mo \rangle ,
\end{equation}
where the superscript \text{I/II} in the dynamical self-energy has been suppressed for convenience. Instead of formulating a BSE for the response function, the ADC approach expresses $M(\omega)$ through\cite{Schirmer1983}
\begin{equation}
    M_{pq}(\omega) = U_p\left[\omega\mathds{1} - K - C\right]^{-1}U^{\dagger}_{q} \;.
\end{equation}
The coupling and interaction matrices can be perturbatively expanded as
\begin{equation}
    \begin{split}
        U_p(n) &= U_p^{(1)} + U_p^{(2)} + \cdots + U_p^{(n)} \\
        C(n)   &= C^{(1)} + C^{(2)} + \cdots + C^{(n)} \;.
    \end{split}
\end{equation}
The ADC formalism then approximates $M(\omega;n)$ systematically by requiring exactness up to the $n$th-order, such that
\begin{equation}
    M_{pq}(\omega;n) = \sum_{\nu = 2}^n M_{pq}^{(\nu)} (\omega) + \mathcal{O}\left(V^{n+1}\right),
\end{equation}
where $M_{pq}^{\nu}$ is the $\nu$th-order contribution in the diagrammatic expansion. In ADC($n$), $U$ and $C$ are expanded to order $n-1$ and $n-2$, respectively.

A direct consequence of this perturbative expansion is the construction of the configuration space. While ADC(2m), ADC(2m+1) and (2m+1,1)-MCDE all restrict the Hilbert space to the lowest (m+1)p-mh and (m+1)h-mp excitations,\cite{Schirmer1983, Riva2025} the hierarchy is built differently between the two methods.
In the (3,1)-MCDE formalism, the configuration space is determined by the order of the underlying Green's function, whereas in ADC the inclusion of higher excitation channels is governed by the perturbation order of the self-energy. Consequently, each (2m+1)-body Green's function in MCDE spans the same excitation space as two successive ADC orders, ADC(2m) and ADC(2m+1) without being equivalent to either. (3,1)-MCDE corresponds to the intermediate ADC(2)-X.\cite{Demartini2026AlgebraicEquation}
\\ \\
Let us now consider the explicit expressions for the effective Hamiltonian blocks arising from the two approaches. Since this work is restricted to the 2p1h and 2h1p sectors, we focus on the (3,1)-MCDE, ADC(2) and ADC(3) approximations. In addition, the well-known ADC(2)-X method is included as an ad-hoc intermediate between ADC(2) and ADC(3).\cite{trofimov1995efficient} Starting from the diagonal matrix elements, all schemes obtain the same diagonal zeroth-order contributions, given by\cite{Schirmer1983, Riva2024}
\begin{equation}
    \begin{split}
        \epsilon_{ij} &= \epsilon_{i}\delta_{ij}, \\
        K_{ijl;mok} &= \left(\epsilon_i + \epsilon_j - \epsilon_l\right) \delta_{im}\delta_{jo}\delta_{lk},
    \end{split}
\end{equation}
where an element $\epsilon_i$ is the HF energy of orbital $i$. Only ADC(3) incorporates the third-order static self-energy correction in the one-body block, reflecting the first non-trivial renormalization of the HF orbital energies.\cite{Schirmer1983} To simplify the expressions between the I and II sectors, we define the occupation projector 
\begin{equation}
    F_{ijk} = (f_i - f_k)(f_j - f_k),
\end{equation}
where $f_i$ is the occupation number, i.e., $f_i = 1$ ($f_i = 0$) for $|k\rangle$ occupied (unoccupied) in the HF ground state $|\Phi_0^N\rangle$. The occupation tensor restricts the blocks to the 2p1h and 2h1p contributions. Then, assuming real orbitals, the modified coupling matrix elements become\cite{Schirmer1991Closed-formMatrices, Riva2024}
\begin{equation}
    \begin{split}
        \text{(3,1)-MCDE:}& \quad U^{\text{I/II}}_{i;mok} = F_{mok}\langle ik||mo\rangle, \\
        \text{ADC(2):}& \quad U^{\text{I/II}}_{i;mok} = F_{mok}\langle ik||mo\rangle, \\
        \text{ADC(2)-X:}& \quad U^{\text{I/II}}_{i;mok} = F_{mok}\langle ik||mo\rangle, \\
        \text{ADC(3):}& \quad U^{\text{I}}_{i;mok}  \\
        & \quad =   F_{mok}\left(\langle ik||mo \rangle + U^{\text{I}, (2)}_{i;mok}\right), \\
                      & \quad U^{\text{II}}_{i;mok}   \\
                      & \quad = F_{mok}\left(\langle ik||mo \rangle + U^{\text{II}, (2)}_{i;mok}\right), \\
    \end{split}
\end{equation}
with
\begin{equation}
\label{eq:u2}
    \begin{split}
        U^{\text{I},(2)}_{i;mok} &= \sum_{pq} \Bigg[ -\frac{1}{2} \frac{\langle mo||pq \rangle \langle ik||pq\rangle}{\epsilon_o + \epsilon_m - \epsilon_p - \epsilon_q}f_p f_q \\
        & \qquad \qquad + \frac{\langle mp||qk\rangle \langle ip||qo \rangle}{\epsilon_q + \epsilon_k - \epsilon_m - \epsilon_p}\bar{f}_p f_q \\
        & \qquad \qquad - \frac{\langle op||qk \rangle \langle ip||qm \rangle}{\epsilon_q + \epsilon_k - \epsilon_o - \epsilon_p}\bar{f}_p f_q\Bigg], \\
        U^{\text{II},(2)}_{i;mok} &= -U^{\text{I},(2)}_{i;mok} \  (f_i \longleftrightarrow \bar{f}_i) \;,
    \end{split}
\end{equation}
and $\bar{f}_i = 1 - f_i$.
The occupation projector compactly describes the 2p1h/2h1p sectors. The only distinction between the methods is therefore the inclusion of the second-order couplings $U^{(2)}$ in ADC(3). Beyond the time-dependent HF (TD-HF) type coupling at first order ($U^{(1)}$), we can identify in  Eq.~\eqref{eq:u2} the particle-particle ladder vertex in the first line, while the term in the second line lives in the particle-hole channel. The term in the third line lives in the transverse particle-hole channel and is the necessary exchange counterpart to the previous term, making $U^{(2)}$ exactly crossing symmetric. In Eq.~\eqref{eq:u2} we also recognize the first-order 2-particle amplitudes familiar from Møller--Plesset perturbation theory,\cite{Szabo2012} highlighting that $U^{(2)}$ effectively acts as a 2-body scattering amplitude dressed by the correlated ground state wave function, approximated at first order.

 At last, we have the modified interaction matrix elements, given by\cite{Schirmer1991Closed-formMatrices, Riva2024}
\begin{equation}
    \begin{split}
        \text{(3,1)-MCDE:}& \quad C^{\text{I/II}}_{ijl;mok} = s_{\text{I/II}}F_{ijl} C^{(1)}_{ijl;mok}F_{mok}, \\
        \text{ADC(2):}& \quad C^{\text{I/II}}_{ijl;mok} = 0\\
        \text{ADC(2)-X:}& \quad C^{\text{I/II}}_{ijl;mok}= s_{\text{I/II}}F_{ijl} C^{(1)}_{ijl;mok}F_{mok} ,  \\
        \text{ADC(3):}& \quad C^{\text{I/II}}_{ijl;mok} = s_{\text{I/II}}F_{ijl} C^{(1)}_{ijl;mok}F_{mok},  
    \end{split}
\end{equation}
with $s_{\text{I}}=+1$ and $s_{\text{II}}=-1$, where
\begin{equation}
\label{eq:c1}
    \begin{split}
        C^{(1)}_{ijl;mok} &= \delta_{lk}\langle ij||mo \rangle  \\
                        & \quad+ \delta_{oj} \langle ik||lm \rangle + \delta_{im} \langle jk||lo \rangle \\
                        & \quad- \delta_{mj}\langle ik||lo \rangle - \delta_{io} \langle jk||lm \rangle 
    \end{split}
\end{equation}
The first term represents the particle-particle ladder and the remaining terms account for the direct particle-hole interactions and their exchange counterparts, which together describe the particle-hole ladders. This distinction is exploited in the screened (3,1)-MCDE (sMCDE)\cite{Romaniello2026DirectEquation} that replaces the bare interaction in all ladders in Eq.~\eqref{eq:c1} with the statically screened one calculated in the (direct) random phase approximation (RPA).

A direct comparison of the effective Hamiltonian blocks reveals that (3,1)-MCDE and ADC(2)-X share the same working equations.\cite{Demartini2026AlgebraicEquation} Whereas ADC(2)-X serves as a manually modified ADC(2), the equivalent Hamiltonian structure arises naturally within the (3,1)-MCDE formalism. Second, ADC(2) differs only by the exclusion of the modified interaction block $C$, while ADC(3) extends the ADC(2)-X/(3,1)-MCDE Hamiltonian by incorporating second-order corrections to both the one-body block and the coupling matrix. 

\subsection{\label{sec:static_theory}Static Self-energy}
Here, we briefly review the static part of the self-energy and refer to the supplemental material for details.
The static self-energy $\Sigma_c(\infty)$ defined in Eq.~\eqref{eq:sigma_infty} can in principle be determined self-consistently in terms of $M(\omega)$ by inserting the Dyson equation Eq.~\eqref{2.6} into its definition. In practice, Eq.~\eqref{2.6} is linearized, 
\begin{equation}
    \label{eq:dyson_linearized}
    G(\omega) \approx G^{0}(\omega) + G^{0}(\omega)\Sigma_c(\omega)G^{0}(\omega) \;.
\end{equation}
Since $\Sigma(\omega)$ depends on the density matrix through its static part, this equation must be determined self-consistently.\cite{vonNiessen1984ComputationalFunction} However, replacing  $\Sigma(\omega)$ in Eq.~\eqref{eq:dyson_linearized} by its dynamic contribution $M(\omega)$ one can show that Eq.~\eqref{eq:sigma_infty} is equivalent to
\begin{equation}
\label{eq:sigma_infty2}
    \Sigma_{c_{pq}}(\infty) = \sum_{rs}\langle pr||qs\rangle \Delta \gamma_{rs} \;,
\end{equation}
where $\Delta \gamma$ is the difference between the correlated and the HF density matrix. Using Rayleigh--Schrödinger perturbation theory it can be expanded as
\begin{equation}
\label{eq:density_expansion}
    \Delta \gamma_{rs} = \Delta \gamma^{(2)}_{rs} + \Delta \gamma^{(3)}_{rs} + \dots \;,
\end{equation}
where the superscripts denote the order in the Coulomb interaction.
This expansion provides a closed expression for the static self-energy that is consistent with the order $n$ of a given level of the ADC approximation.
For instance, using $\Delta \gamma^{(2)}$ in Eq.~\eqref{eq:sigma_infty2} gives the static self-energy to third order, often referred to as $\Sigma(3)$ in the ADC literature.\cite{Trofimov2005MolecularApproach} $\Delta \gamma^{(2)}$ is the so-called unrelaxed MP2 density matrix,\cite{Amos1980CorrectionsTheory} that does not take into account the static density response induced by the dynamical self-energy, i.e. orbital relaxation. Orbital relaxation is introduced by using $\Sigma(\omega)$ in Eq.~\eqref{eq:dyson_linearized}, which allows the static part of the self-energy to relax due to the dynamical self-energy. Eq.~\eqref{eq:dyson_linearized} can be transformed into a linear eigenproblem\cite{vonNiessen1984ComputationalFunction} whose solution, at second order, gives the relaxed density matrix that can be used to calculate analytical MP2 gradients.\cite{Salter1987TheoryApproach, Salter1989AnalyticDerivatives} The corresponding self-energy is termed $\Sigma(3+)$.\cite{Trofimov2005MolecularApproach} 

A third-order approximation to $\Sigma_c(\infty)$ would be consistent with the ADC(3) approximation. However, Schirmer and co-workers observed $\Sigma(3)$ to be a poor approximation to the static self-energy and evaluated Eq.~\eqref{eq:density_expansion} to third order in the Coulomb interaction, giving the $\Sigma(4)$ approximation. Also $\Sigma(4)$ approximates the static self-energy rather poorly\cite{Trofimov2005MolecularApproach}, but the $\Sigma(4+)$ approximation, that, in analogy to $\Sigma(3+)$, takes into account orbital relaxation effects, has been found to be a good approximation to $\Sigma_c(\infty)$.\cite{Trofimov2005MolecularApproach} We note that the orbital-relaxed MP3 density matrix is not equivalent to the fully relaxed MP3 density matrix since the latter also includes the effect of amplitude relaxation. In the following, we will refer to the ADC with $\Sigma_c(\infty)=0$ as $\Sigma(1)$, to indicate that the static self-energy is complete to first order in the Coulomb interaction.

\subsection{Non-Dyson approximation}

While the Dyson formalism yields both ionization potentials and electron affinities from a single eigenvalue problem, practical applications often require mainly one part of the spectrum. In such cases, the nD formalism provides an alternative approach by separating the EA and IP sectors, and reducing the computational cost.\cite{Schirmer1998APropagator} Starting from the general effective Hamiltonian \eqref{2,1:effectiveHamiltonian}, we can project out the 2p1h sector, and obtain
\begin{equation}
    H^{\text{II}}_{\text{eff}}(\omega) = 
        \begin{pNiceArray}{c|ccc}
        \\
        \epsilon + \Sigma_c(\infty) + M^{>}(\omega)& &  U^{\text{II}} & \\ & \\
        \hline \\ 
        \left(U^{\text{II}}\right)^{\dagger} & & {K^{\text{II}}} + {C^{\text{II}}}  \\ \\ 
        \end{pNiceArray},
\end{equation}
where $M^{>}$ is defined as
\begin{equation}
    M_{pq}^{>}(\omega) = U_p^{\text{I}}\left[\mathds{1}\omega - (K^I + C^I)\right]^{-1}U^{\text{I}^{\dagger}}_q.
\end{equation}
Evaluating $M^{>}(\omega)$ at the HF QP energies, together with explicit symmetrization, we obtain the nD effective Hamiltonian of the IP sector
\begin{equation}
    H^{\text{nD}}_{\text{eff}} = 
        \begin{pNiceArray}{c|ccc}
        \\
        \epsilon + \widetilde{\Sigma}_c(\infty)& &  U^{\text{II}} & \\ & \\
        \hline \\ 
        \left(U^{\text{II}}\right)^{\dagger} & & {K^{\text{II}}} + {C^{\text{II}}}  \\ \\ 
        \end{pNiceArray},
\end{equation}
where the modified static self-energy reads
\begin{equation}
\label{eq:nD-sigma}
    \begin{split}
        \widetilde{\Sigma}_{c_{pq}}(\infty) &= \Sigma_{c_{pq}}(\infty) \\
        &\qquad + \frac{1}{2}\left[M^>_{pq}(\epsilon_p) + M^>_{pq}(\epsilon_q)\right].
    \end{split}
\end{equation}
This type of symmetrization is also the standard common choice in QP approximations to the $GW$ self-energy,\cite{Kotani2007, Sheng2022, Romanova2023} but a symmetrization of the form $M^>_{pq}(\frac{1}{2}(\epsilon_p + \epsilon_q))$ is a valid choice as well.\cite{Loos2026FromApproximations}
Similarly, the EA sector is retrieved by projecting out the 2h1p sector. 
\subsection{Davidson}
Direct diagonalization of the effective Hamiltonians is prohibitively expensive for medium- and large-sized systems due to its $\mathcal{O}(N^9)$ scaling with system size.\cite{Riva2023} The standard Davidson iterative solver obtains the $n$ lowest eigenvalues of the system.\cite{Davidson1975} However, one is often more interested in specific eigenvalues. An alternative to the standard Davidson routine is known as harmonic Davidson, in which one specifies an energy interval (or shift),\cite{morgan1991computing, zuev2015new, Huang2022TwoPair} and the method preferentially converges eigenvalues in that region. The target shift is chosen as the midpoint of the interval, $\sigma = \frac{E_{max} + E_{min}}{2}$. Harmonic Davidson then preferentially converges to the interior eigenvalues located closest to this shift. Although the restriction decreases the computational time, two problems remain. Firstly, the number of eigenvalues within an energy window is uncertain, and the harmonic Davidson solver expects it as an initial parameter. Secondly, it does not target individual eigenvalues, which could particularly be useful when considering ionization potentials. A small modification of the (harmonic) Davidson method fixes these complications, by simply targeting a single energy eigenvalue $\sigma$ chosen from an approximated energy value beforehand. Our linearization approximation provides an inexpensive estimate of the target energy $\sigma$, which is subsequently used as the shift for the IP-target Davidson algorithm. The Davidson procedure then refines this estimate to the converged QP energy. The linearization procedure starts from the downfolded Dyson equation, given by
\begin{equation}
    G(\omega) = \left[\mathds{1}\omega - \epsilon^{\text{1p}} - M(\omega)\right]^{-1},
\end{equation}
where the 1-body energy $\epsilon^{\text{1p}}$ and frequency-dependent correlated self-energy $M(\omega)$ are given by
\begin{equation}
    \begin{split}
        \epsilon^{1p} &= \epsilon + \Sigma_c(\infty), \\
        M(\omega) &= U[\mathds{1}\omega - (K + C)]^{-1}U^{\dagger},
    \end{split}
\end{equation}
and $\epsilon$ is the matrix of HF energies.
The full spectrum corresponds to all poles of the Green's function. Starting from the excitation energies,
\begin{equation}
    E_i = \epsilon^{\text{1p}}_i + M_i(E_i) ,
    \label{dav:excitation energy}
\end{equation}
where $M_i$ denotes a diagonal element of $M$, we approximate $M(E_i)$ by Taylor expanding around the HF orbital energy, such that
\begin{equation}
    \begin{split}
    M(E_i)
        &= M(\omega)\Big|_{\omega=\epsilon_i}
        + (E_i-\epsilon_i)\frac{\partial M(\omega)}{\partial\omega}\Big|_{\omega=\epsilon_i} \\
        & \qquad \qquad \quad \ \ + \mathcal{O}\!\left((E_i-\epsilon_i)^2\right) \\
        &\approx M(\epsilon_i)
        + (E_i-\epsilon_i)\frac{\partial M(\omega)}{\partial\omega}\Big|_{\omega=\epsilon_i},
    \end{split}
\end{equation}
where higher-order terms are neglected, resulting in the first-order linearization approximation. By recombining the different terms in \eqref{dav:excitation energy} a closed form of the excitation energy can be found as
\begin{equation}
    \begin{split}
        E_i &\approx \epsilon_i + Z_i\left[\Sigma_{c_i}(\infty) + M_i(\epsilon_i)\right] \\
        &=  \epsilon_i + Z_i\Sigma_{c_i}(\epsilon_i),
    \end{split}
\end{equation}
with the Lorentzian weight or $Z$-factor given by
\begin{equation}
    Z_i = \left[1 - \frac{\partial M_i(\omega)}{\partial \omega}\Big|_{\omega = \epsilon_i}\right]^{-1}.
\end{equation}
In the explicit expression for the dynamical self-energy, the inverse of the three-body part still requires diagonalization. Defining $K + C = VDV^{-1}$, together with the identity $(ABC)^{-1} = C^{-1}B^{-1}A^{-1}$, the linearized excitation energy can be written as
\begin{equation}
    E_i \approx \epsilon_i + Z_i \left[\Sigma_{c_i}(\infty) + \sum_p \frac{|W_{i;p}|^2}{\epsilon_i - d_p}\right],
\end{equation}
where $W_{i;p} = \sum_{\nu} V^{-1}_{p;\nu}U_{\nu;i}$ are the residues of the dynamical self-energy ($\nu$ runs over the three-body states). Both the excitation energy and the residues depend explicitly on the 3-body eigenvalues. To avoid diagonalization of a large matrix, we approximate $K+C$ as diagonal, so that $V = 1$, $W \rightarrow U$, giving
\begin{equation}
    E_i \approx \epsilon_i + \widetilde{Z}_i\left[\Sigma_{c{_i}}(\infty) + \sum_p \frac{|U_{i;p}|^2}{\epsilon_i - (k + c)_{pp}}\right],
\end{equation}
and 
\begin{equation}
    \widetilde{Z}_i = \left[1 + \sum_p \frac{|U_{i;p}|^2}{(\epsilon_i - (k+c)_{pp})^2}\right]^{-1},
\end{equation}
where $(k+c)_{pp}$ are the diagonal 3-body elements. Neglecting the off-diagonal couplings in the 3-body block might shift the pole positions significantly, making the approximation unsuitable as a stand-alone method for larger systems.
While an independent implementation of the linearization approximation results in large errors with respect to the full spectrum, applying it as a precursor followed by the Davidson method converges the approximate value to the true eigenvalue. Concretely, the IP-target Davidson solves the shifted inverse operation defined as\cite{Huang2022TwoPair}
\begin{equation}
\label{eq:inverse}
    (H_{\text{eff}} - \mathds{1}\sigma_{\text{IP}})^{-1} x = \mu x,
\end{equation}
where $\sigma_{\text{IP}}$ is the linearized ionization potential, and the shifted eigenvalue $\mu$ is defined by
\begin{equation}
    \mu = \frac{1}{\lambda - \sigma_{\text{IP}}}.
\end{equation}
The eigenvalues close to $\sigma_{\text{IP}}$ thus become dominant eigenvalues of the shifted problem.
As forming the explicit inverse in Eq.~\eqref{eq:inverse} is prohibitive, we impose a harmonic Petrov–Galerkin condition with test space $(H_{\text{eff}} - \mathds{1}\sigma_{\text{IP}})V$ instead, which targets the same eigenvalues using only matrix–vector products with $H_{\text{eff}}$.\cite{morgan1991computing, paige1995approximate, zuev2015new}

In contrast to the standard or harmonic Davidson algorithm, the present approach focuses on a single state without attempting to converge all n-lowest eigenvalues or all eigenvalues within an interval. Therefore, the dependence on the preferred number of eigenvalues and the specific energy window vanishes. Furthermore, to stabilize the selection procedure, the IP-target Davidson routine is initialized with the dominant one-particle configuration corresponding to the occupied orbital of interest. After solving the projected harmonic eigenvalue problem, the resulting Ritz vectors are filtered according to the square roots of the QP weights,
\begin{equation}
    \Omega_k = \sqrt{\sum_{i \in p} |x_{k,i}|^2} ,
\end{equation}
where \(x_{k,i}\) denotes the one-particle component of the \(k\)th Ritz vector. The Ritz vector maximizing \(\Omega_k\) is then selected for the subsequent Davidson iteration. The algorithm thereby continuously tracks the desired QP state, even in the vicinity of satellite solutions, obtaining the desired ionization potential.

\section{Computational Details}
We have implemented the ADC(2), ADC(2)-X/(3,1)-MCDE, and ADC(3) approximations into the BAND engine of a locally modified version of the Amsterdam modelling suite (AMS)\cite{Baerends2025} as well as a stand-alone package built on top of pySCF.\cite{Sun2020, Sun2026TheProject} Both implementations make use of density fitting (for the implementation of DF in AMS see Ref.~\citenum{Spadetto2023}) and implement the root-following Davidson algorithm described above. The correctness of the implementations has been verified by comparison to the Dyson-ADC(3) results in Ref.~\citenum{Marie2026AnSelf-Energy} as well as against pySCF for nD-ADC.

Coupled cluster with singles and doubles (CCSD) density matrices have been calculated using pySCF. The code to evaluate the unrelaxed MP2 and MP3 density matrices has been generated using the p$^{\dagger}$q-package.\cite{Liebenthal2025AutomatedPackage} Orbital relaxation has been taken into account using the method of Schirmer and co-workers,\cite{vonNiessen1984ComputationalFunction} (See also Ref.~\citenum{Bruneval2026GWEquation}) which is equivalent to the $Z$-vector approach for MP2.\cite{Handy1984OnFunctions} The triple amplitudes in the MP3 density matrices have been factorized by Laplace transforming the energy denominators\cite{Almlof1991, Haser1992} using minimax grids\cite{Takatsuka2008} as implemented in the GreenX library.\cite{Azizi2023Time-frequencyCalculations} Detailed working equations and derivations are provided in the supporting information. All calculations have been performed with six Laplace points, which is known to suffice to converge MP2 energy denominators in systems with large HOMO-LUMO gap (like the ones we consider in this work) to meV precision.\cite{Doser2009a}

\section{\label{sec:results}Results}

\begin{figure*}[hbt!]
    \centering
    \includegraphics[width=\linewidth]{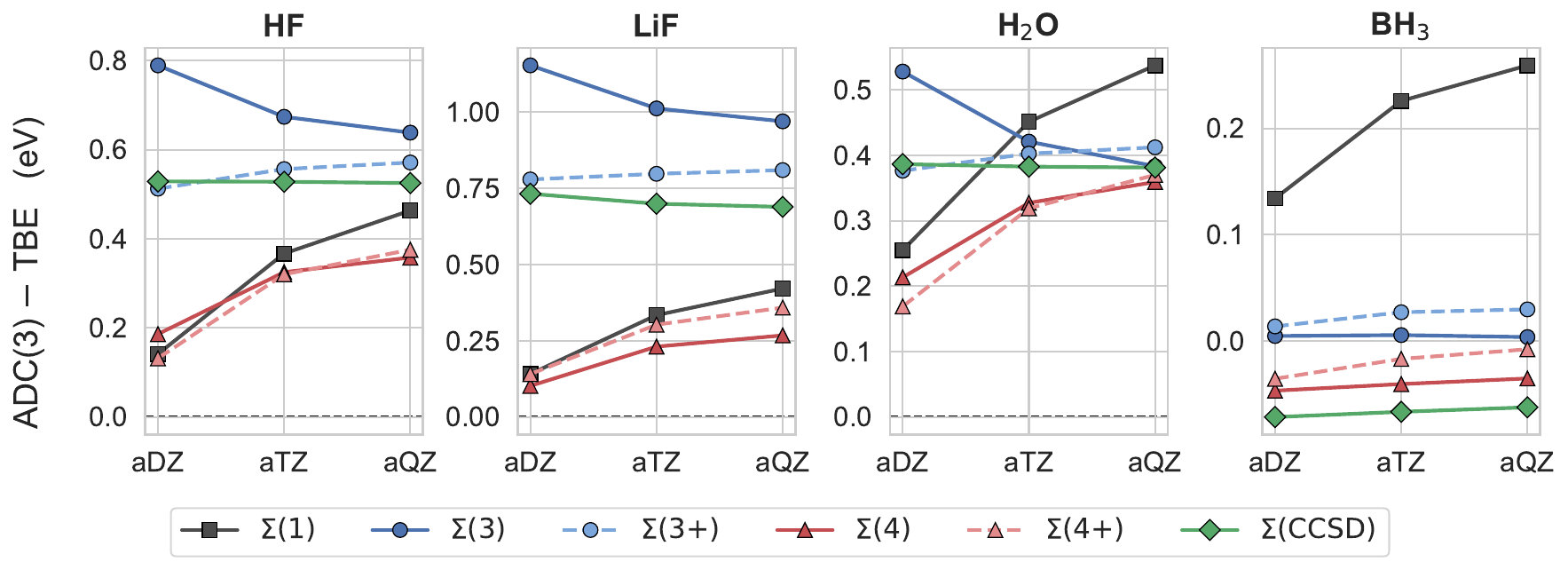}
    \caption{Errors of ADC(3) with different approximations to $\Sigma_c(\infty)$ to the TBE calculated at different basis sets, $\text{IP}^{\text{ADC(3)}}[\text{aXZ}] - \text{IP}^{\text{TBE}}[\text{aXZ}]$, for $X = D,T,Q$. All values are in eV.}
    \label{fig:basis_convergence_error_vs_tbe_state1}
\end{figure*}

We now turn to the discussion of our numerical results for a benchmark set for which near-full configuration interaction (CI) results using a selected CI (sCI) algorithm\cite{Huron1973, Garniron2018SelectedPerturbation, Garniron2019QuantumPrograms} have been calculated recently [referred to as theoretical best estimates (TBE) in the following].\cite{Marie2024} The set comprises 58 ionization potentials of 23 small molecules. In the following results obtained with the aug-cc-pVQZ basis set, we only show 56 out of the 58 IPs, because CO$_2$ misses 2 sCI reference values at this basis set. Furthermore, our nD-ADC(2)-X calculation for CH$_2$O only converged onto 3 physical QP roots, instead of 4. Therefore, for nD-ADC(2)-X, we also omit the 4th IP for this system in the following comparisons. This is due to a shortcoming of the nD formalism, which folds the coupling to the 2p1h sector into a static correction, $\widetilde{\Sigma}_c(\infty)$ evaluated at the reference orbital energy (See Eq.~\ref{eq:nD-sigma}). This is a QP approximation that is only valid when the true pole lies close to the HF reference orbital energy, and tends to break down in the semi-valence and semi-core regions, where several near-degenerate 2h1p configurations mix strongly with the nominal 1h state.\cite{Cederbaum1975, Cederbaum1977}

\subsection{Basis set convergence}

To prepare our comparison of the various ADC approximations against the reference values of Ref.~\citenum{Marie2024}, we first discuss how the choice of basis set used will affect this comparison. To this end, Fig.~\ref{fig:basis_convergence_error_vs_tbe_state1} shows for four representative molecules the errors of ADC(3) with different approximations to $\Sigma_c(\infty)$ to the TBE calculated at different basis sets, $\text{IP}^{\text{ADC(3)}}[\text{aXZ}] - \text{IP}^{\text{TBE}}[\text{aXZ}]$, for $X = D,T,Q$. Clearly, the errors generally change substantially when the basis set is varied. However, this behaviour is molecule-dependent. For HF, LiF, and H$_2$O, the dependence is more substantial than for BH$_3$, because the former three molecules' HOMO is localized on a very electronegative element, linked to slow basis set convergence.\cite{Bruneval2020, Baum2026PredictingCalculations} The same conclusions can be drawn for ADC(2)-X which we do not show here.

The density matrix used to evaluate the static self-energy has a major influence on the basis set convergence. An increasing curve indicates that the respective method's basis set convergence is slower than that of the sCI TBE, while a decreasing curve signifies a faster basis set convergence. Consequently, the data show that the basis set convergence of ADC(3)-$\Sigma(1)$ ($\Sigma_c(\infty)=0$) is slowest, followed by ADC(3)-$\Sigma(4)$. The curves for ADC(3)-$\Sigma(\text{CCSD})$ and also ADC(3)-$\Sigma(3+)$ are uniformly almost flat, while ADC(3)-$\Sigma(3)$ shows a very fast basis set convergence.

For the slowly converging systems in Fig.~\ref{fig:basis_convergence_error_vs_tbe_state1}, the spread between the different approximations to $\Sigma_c(\infty)$ decreases from 0.69 eV at aug-cc-pVDZ to 0.39 eV at aug-cc-pVQZ, so that a comparison against the TBE in a small basis set would be skewed by basis set incompleteness rather than by the quality of $\Sigma_c(\infty)$ itself. For this reason, all conclusions about the accuracy of the different ADC methods in Sec.~\ref{sec:adc_tbe} will be based on the aug-cc-pVQZ basis set. Extrapolating the observed aDZ $\to$ aTZ $\to$ aQZ convergence, the residual basis set incompleteness of the mean error at aug-cc-pVQZ amounts to 0.05 eV for ADC(3)-$\Sigma(1)$ and to less than 0.015 eV for all correlated densities, which sets a very small error bar for our comparison.

\subsection{\label{sec:static}Comparison of static self-energies}

\begin{figure}[hbt!]
    \centering
    \includegraphics[width=\linewidth]{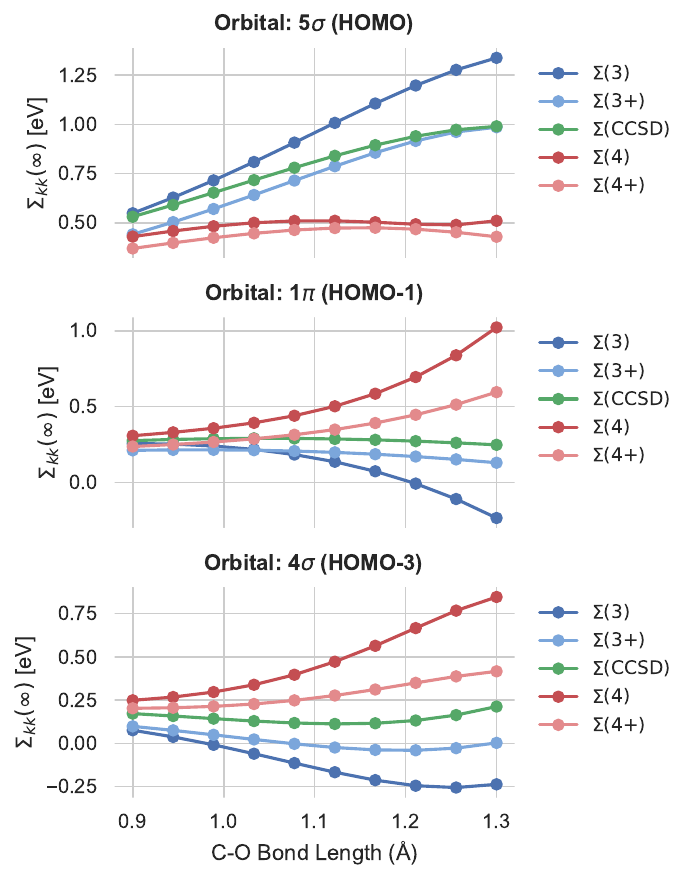}
    \caption{Diagonal element of the static self-energy contribution to the CO ionization potential as a function of bond length, for the $5\sigma$, $1\pi$, and $4\sigma$ orbitals, using different approximations to $\Sigma_c(\infty)$ and the aug-cc-pVQZ basis set. All values are in eV.}
    \label{fig:adc3_co_scan_plot_qz}
\end{figure}

Next, we compare different approximations to $\Sigma_c(\infty)$. Following \citet{Trofimov2005MolecularApproach}, in Fig.~\ref{fig:adc3_co_scan_plot_qz} we plot different diagonal elements of $\Sigma_c(\infty)$ evaluated with different approximations to $\Delta \gamma$ for CO as a function of bond length,\cite{Trofimov2005MolecularApproach} which directly probes the quality of the density matrix at different correlation strengths, in the aug-cc-pVQZ basis set. We find that for all three states, $\Sigma(3+)$ closely tracks the CCSD reference, while especially for the 5$\sigma$ state, $\Sigma(4+)$ shows substantially larger deviations. In agreement with Ref.~\citenum{Trofimov2005MolecularApproach}, orbital relaxation in both cases substantially improves agreement with the reference curve.

\subsection{Dyson vs non-Dyson}

\begin{figure*}[hbt!]
    \centering
    \includegraphics[width=\linewidth]{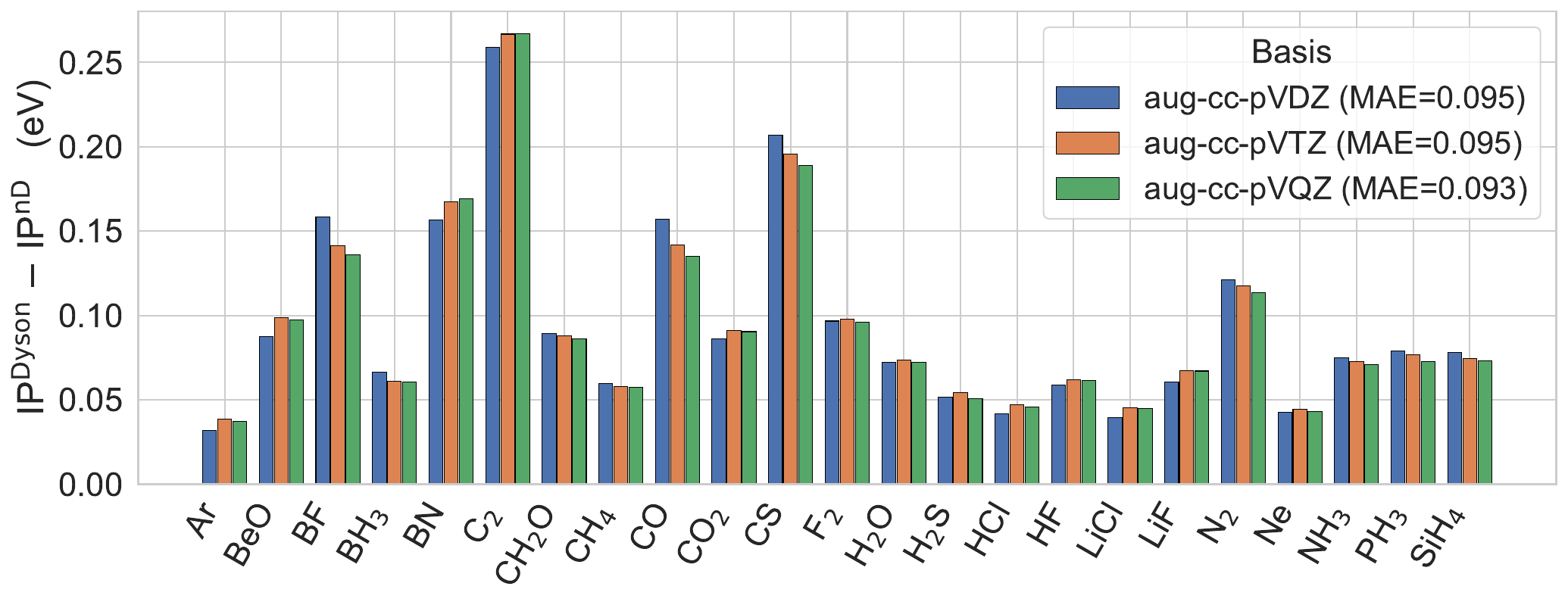}
    \caption{Deviations of the first IPs of the 23 molecules in the Marie--Loos set between Dyson-ADC(3) and non-Dyson (nD)-ADC(3) for three different basis sets. Mean absolute errors (MAE) are shown in the legend. All values are in eV.}
    \label{fig:dyson_vs_nondyson_adc3_histogram_homo}
\end{figure*}

We now turn to the comparison of Dyson- and nD-ADC(3). The most systematic comparison of these methods has been reported by \citet{Trofimov2005MolecularApproach} in 2005, who concluded that the nD approximation has only a small, negligible effect on the accuracy of the ADC scheme, while other factors, most notably the density matrix used to evaluate $\Sigma_c(\infty)$ plays a much larger role. This work has been very influential in the sense that most of the current ADC work exclusively focuses on the nD formalism, often citing \citet{Trofimov2005MolecularApproach} to justify this approximation. Very recently, \citet{Marie2026AnSelf-Energy} have implemented Dyson-ADC(3) and compared their results against the nD-ADC(3) implementation in pySCF,\cite{Banerjee2019Third-orderImplementation} finding rather large discrepancies between both methods. However, theycompared Dyson-ADC(3)-$\Sigma(1)$ against nD-ADC(3)-$\Sigma(3)$, and it is well-known (we will also investigate these aspects in some detail in Sec.~\ref{sec:adc_tbe}) that the static approximation has a large effect on the accuracy of an ADC calculation.\cite{Trofimov2005MolecularApproach}

In Fig.~\ref{fig:dyson_vs_nondyson_adc3_histogram_homo} we compare Dyson-ADC(3)-$\Sigma(3)$ against nD-ADC(3)-$\Sigma(3)$  as implemented in pySCF for the HOMO energy level for all 23 molecules in the benchmark set of \citet{Marie2024}. Since all values have been calculated with the same static density matrix, this comparison isolates the nD-approximation as the only source of discrepancy. We find, independent of the basis set used, that nD- and Dyson-ADC(3)-$\Sigma(3)$  differ substantially. The error of the nD approximation is of the order of 0.09-0.10 eV on average. Furthermore, for CS, the difference is of the order of 0.2 eV, and for the strongly correlated C$_2$ molecule, the deviation between both methods exceeds 0.25 eV. In light of these substantial deviations that are of the order of the often cited "spectroscopic accuracy", we conclude that the error introduced by the nD-approximation is substantial and not negligible as claimed by \citet{Trofimov2005MolecularApproach}. 

\begin{table}[htbp]
\centering
\caption{Deviation between Dyson- and nD-ADC(3), resolved by ionization order, using the aug-cc-pVQZ basis set. $N$ is the number of IPs in each category. $\overline{|\Delta|}$ and $\Delta_{\max}$ are the mean and maximum absolute errors between the two methods. All values are in eV.}
\label{tab:dyson_nd_depth}
\begin{tabular}{c c c c }
\hline\hline
IP order & $N$ & $\overline{|\Delta|}$ & $\Delta_{\max}$  \\
\hline
1st  & 23 & 0.093 & 0.267 \\
2nd  & 22 & 0.085 & 0.288 \\
3rd+ & 11 & 0.127 & 0.262 \\
\hline\hline
\end{tabular}
\end{table}

The QP-type approximation inherent to the nD approximation is most reliable for poles close to the HF orbital energies, and thus becomes harder to justify for semi-valence and semi-core states. Table~\ref{tab:dyson_nd_depth} quantifies the nD error for these deeper states as well. It shows that the discrepancies found for the frontier IPs also persist for deeper IPs, even though the deviation between the nD and Dyson formalisms does not increase uniformly with IP depth.

\subsection{\label{sec:adc_tbe}Accuracy against TBE}

\begin{figure}[hbt!]
    \centering
    \includegraphics[width=\linewidth]{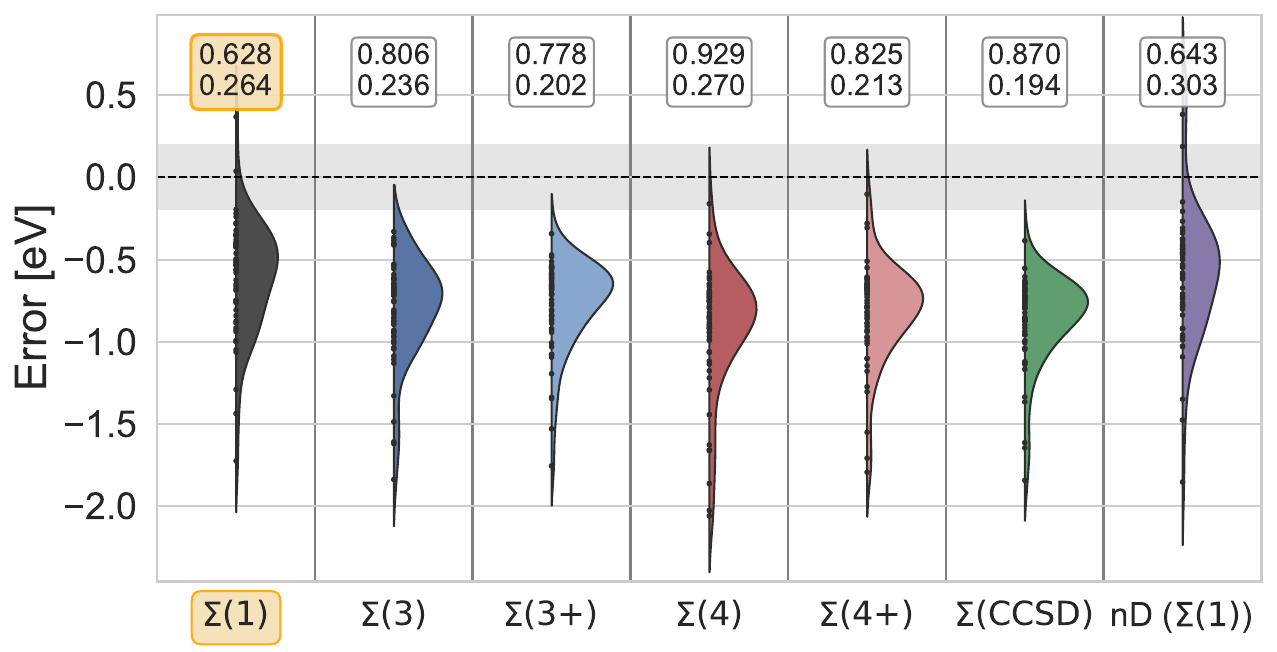}
    \caption{Error distributions of the ADC(2)-X approximation with different approximations to $\Sigma_c(\infty)$ against the sCI TBEs using the aug-cc-pVQZ basis set. The first number over each violin denotes the MAE, the second $\sigma^{\text{MAE}}$. All values are in eV.}
    \label{fig:adc2x_vs_tbe_violin_bothstates_qz}
\end{figure}

We now compare the accuracy of the ADC(2)-X and ADC(3) approximations against the TBEs. The results for ADC(2)-X for the different approximations to $\Sigma_c(\infty)$ are shown in Fig.~\ref{fig:adc2x_vs_tbe_violin_bothstates_qz}, together with mean absolute errors (MAE) as well as $\sigma^{\textrm{MAE}} = \frac{1}{N}\sum_i|\Delta_i - \bar{\Delta}|$, with $\Delta_i$ the error of the $i$th sample and $\bar{\Delta}$ the mean signed error. This quantity represents the mean absolute error about the mean, that eliminates a constant bias.
This figure also shows the results for nD-ADC(2)-X as implemented in pySCF.\cite{Banerjee2019Third-orderImplementation} For the sake of this comparison, we emphasize that this implementation evaluates $\Sigma_c(\infty)$ using $\gamma^{\text{HF}}$, and therefore, the impact of the nD approximation on the MAE can be directly verified through comparison against the left-most violin in Fig.~\ref{fig:adc2x_vs_tbe_violin_bothstates_qz} that shows the results for Dyson-ADC(2)-X-$\Sigma(1)$. Both variants give almost identical MAEs. Generally, all ADC(2)-X approximations perform rather poorly, significantly underestimating the reference IPs. This tendency has been pointed out a long time ago by Cederbaum.\cite{Cederbaum1978CorrelationHydrocarbons} With a MAE of 0.628 eV, ADC(2)-X-$\Sigma(1)$ is the best-performing variant. Errors are comparable to those of the CC2 approximation.\cite{Marie2024} Improving the static part of the self-energy worsens the results. When a very accurate density is used as in ADC(2)-X-$\Sigma(\textrm{CCSD})$, the MAE increases to 0.870 eV.\cite{Paggi2026CoreEquation} 
 
It is however to be noted that improving the static part of the ADC(2)-X self-energy leads to a visibly narrower error distribution, as indicated by the $\sigma^{\mathrm{MAE}}$ values in Fig.~\ref{fig:adc2x_vs_tbe_violin_bothstates_qz}. ADC(2)-X-$\Sigma(\textrm{CCSD})$ has a very small value of $\sigma^{\mathrm{MAE}} = 0.194$~eV at the aQZ level, which shows that almost 700 meV of the MAE of 0.870 eV is a systematic shift. Even for ADC(2)-X-$\Sigma(1)$, the systematic bias approaches 400 meV. Out of all methods shown in Fig.~\ref{fig:adc2x_vs_tbe_violin_bothstates_qz}, nD-ADC(2)-X has the largest value of $\sigma^{\mathrm{MAE}}$. Comparison of $\Sigma(3)$ to $\Sigma(3+)$ and of $\Sigma(4)$ to $\Sigma(4+)$ also reveals that orbital relaxation in $\Sigma_c(\infty)$ always reduces $\sigma^{\mathrm{MAE}}$.

\begin{figure}[hbt!]
    \centering
    \includegraphics[width=\linewidth]{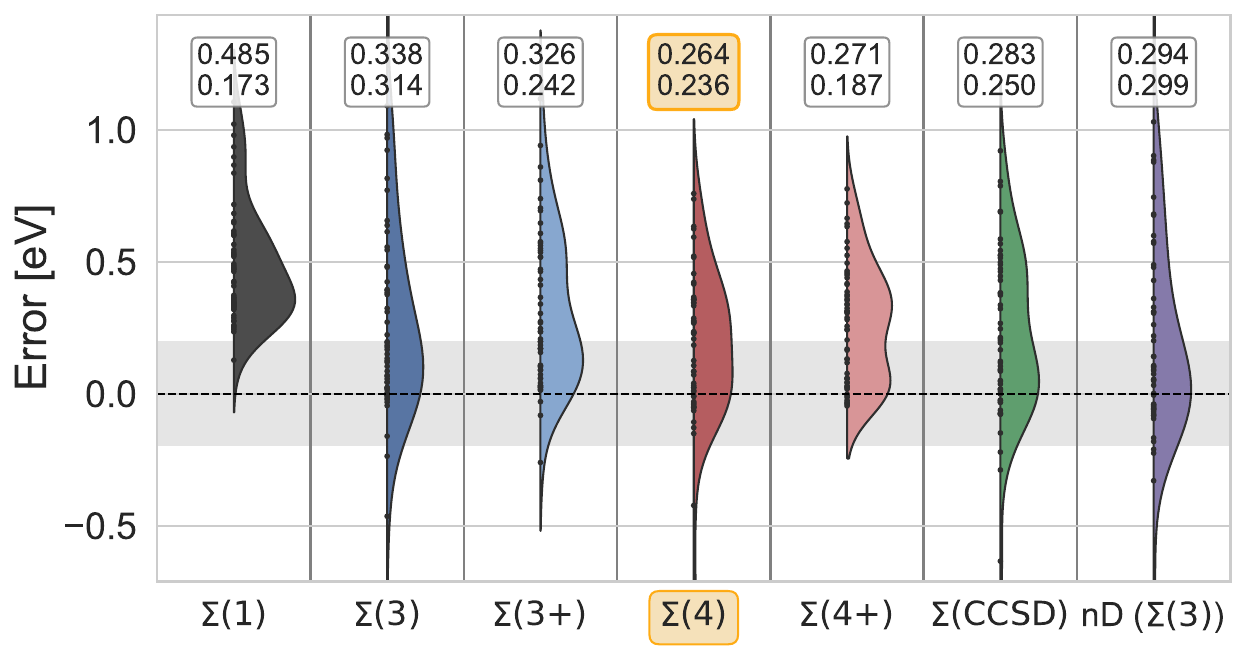}
    \caption{Error distributions of the ADC(3) approximation with different approximations to $\Sigma_c(\infty)$ against the sCI TBEs using the aug-cc-pVQZ basis set. The first number over each violin denotes the MAE, the second $\sigma^{\text{MAE}}$. All values are in eV.}
    \label{fig:dyson_adc3_vs_tbe_violin_bothstates_qz}
\end{figure}

We now turn to the ADC(3) results whose error statistics are shown in Fig.~\ref{fig:dyson_adc3_vs_tbe_violin_bothstates_qz} together with the corresponding MAEs and $\sigma^{\mathrm{MAE}}$ values. Compared to ADC(2)-X, the MAEs are much smaller, with the best-performing variant ADC(3)-$\Sigma(4)$ yielding a MAE of 0.264 eV. In conclusion, Dyson-ADC(3) significantly outperforms $G_0W_0$@HF or CC2,\cite{Marie2024} but it does not reach the accuracy of EOM-CCSD\cite{Marie2024}, state-of-the-art vertex-corrected $GW$-based methods,\cite{Tolle2026ConnectionCluster, Pavlyukh2026ApproachingSelf-Energies} or $G_0W_0$ with static self-energy evaluated with a correlated density matrix. This is consistent with previous ADC(3) benchmarks.\cite{Opoku2025a, Banerjee2023AlgebraicSpectra} Interestingly, compared to ADC(2)-X, ADC(3) does mostly eliminate the constant bias, and slightly overshoots the TBE reference now. However, the $\sigma^{\mathrm{MAE}}$ improve only marginally and even worsen for the $\Sigma(3)$ and $\Sigma(3+)$ variants compared to ADC(2)-X. ADC(3)-$\Sigma(1)$ that evaluates $\Sigma_c(\infty)$ with the HF density matrix has the lowest $\sigma^{\mathrm{MAE}}$ out of all ADC(3) methods. As previously observed for ADC(2)-X, also for ADC(3) the orbital-relaxed $\Sigma(3+)$ and $\Sigma(4+)$ approximations to $\Sigma_c(\infty)$ lead to narrower error distributions compared to their $\Sigma(3)$ and $\Sigma(4)$ counterparts.

Finally, we comment again on the basis set effect. \citet{Marie2026AnSelf-Energy} found an MAE of Dyson-ADC(3)-$\Sigma(1)$ for the exact same IPs of 0.314 eV, using the aug-cc-pVDZ basis set, consistent with our own calculations. The difference to the 0.485 eV we find in this work is entirely a basis set effect. With 0.217 eV, also Dyson-ADC(3)-$\Sigma(4+)$/aug-cc-pVDZ gives a significantly better MAE against sCI/aug-cc-pVDZ, compared to 0.271~eV we obtain when performing the comparison at aug-cc-pVQZ.

\subsection{\label{sec:sMCDE}Screened (3,1)-MCDE}

\begin{figure}
    \centering
    \includegraphics[width=\linewidth]{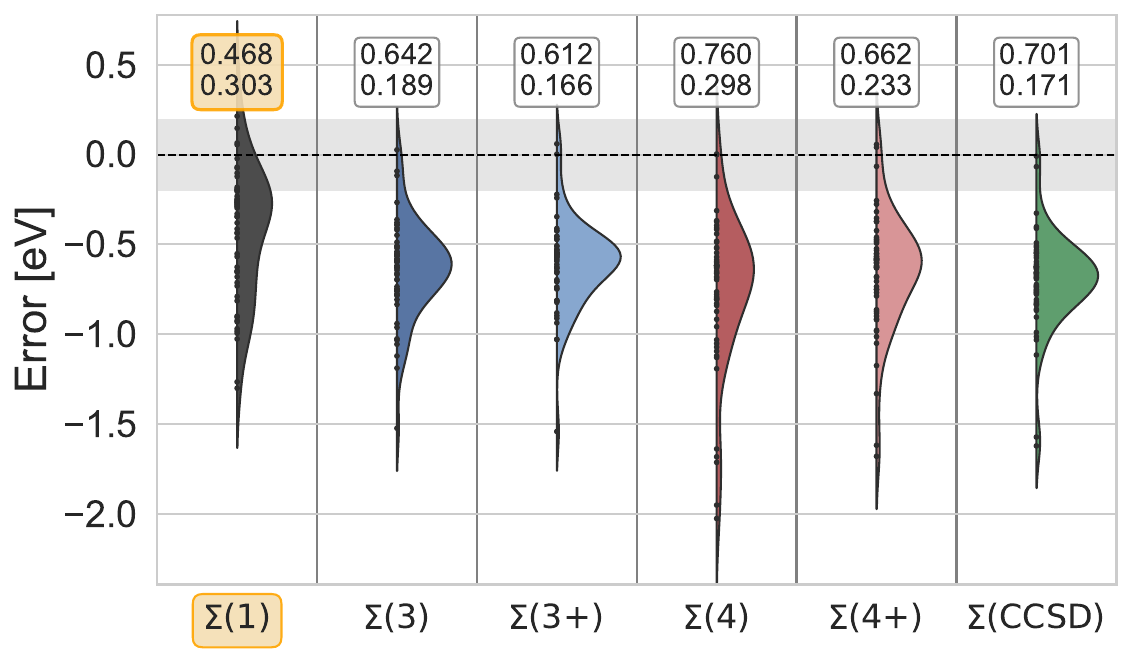}
    \caption{Error distributions of the screened (3,1)-MCDE  approximation with different approximations to $\Sigma_c(\infty)$ against the sCI TBEs using the aug-cc-pVQZ basis set. The first number over each violin denotes the MAE, the second $\sigma^{\text{MAE}}$.  All values are in eV.}
    \label{fig:screened_adc2x_vs_tbe_violin_bothstates_qz}
\end{figure}

The recently introduced screened (3,1)-MCDE\cite{Romaniello2026DirectEquation} replaces all ladder interactions in the standard (3,1)-MCDE with the statically screened Coulomb interaction. More precisely, this is achieved by replacing the HF kernel in the particle-hole channel with the statically screened $GW$ kernel and the $T$-matrix ladders with statically screened ladders.\cite{Romaniello2012, Joost2022a} Benchmark results using the different approximations to $\Sigma_c(\infty)$ calculated with aug-cc-pVQZ are shown in Figure~\ref{fig:screened_adc2x_vs_tbe_violin_bothstates_qz}. As expected, this removes some of the systematic underestimation of ADC(2)-X, since screening suppresses the electron-hole interactions that otherwise lead to underbinding in molecules.\cite{Forster2024} 
Screening also improves $\sigma^{\mathrm{MAE}}$ for the most accurate $\Sigma(3)$, $\Sigma(3+)$ and $\Sigma(\textrm{CCSD})$ variants, even slightly outperforming the best ADC(3) variant.

\subsection{Accuracy for first, second, and third ionization potentials}

\begin{figure}[hbt!]
    \centering
    \includegraphics[width=\linewidth]{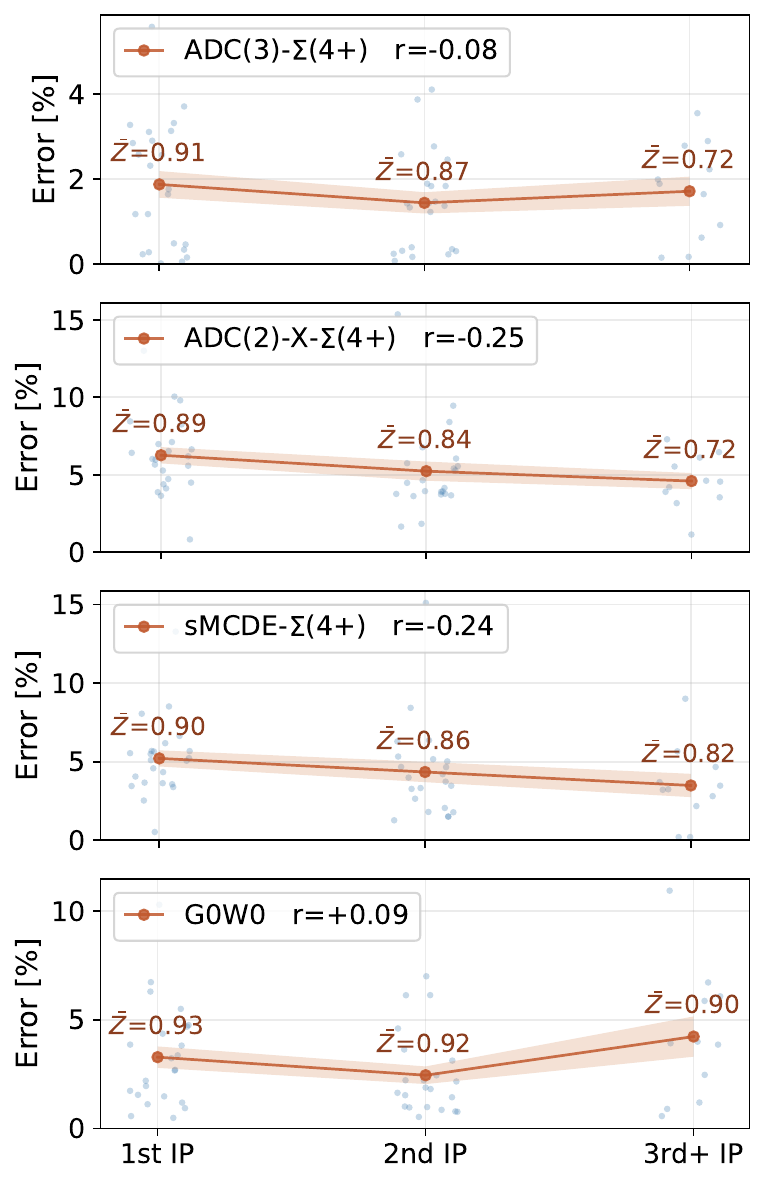}
    \caption{Relative IP error (in \%) against the sCI TBEs, grouped by ionization order, for ADC(3), ADC(2)-X, screened (3,1)-MCDE, using $\Sigma(4{+})$, and G0W0 (Ref.~\citenum{Marie2024}) all in the aug-cc-pVQZ basis set.}
    \label{fig:z_vs_error_qz}
\end{figure}

So far, we have only discussed the performance of these methods for the aggregated set of IPs. However, it is often important to distinguish between frontier IPs and semi-valence states. The former are usually characterized by large QP weights, while semi-valence and semi-core ionizations are often accompanied by shake-up satellite transitions, leading to a noticeable decrease of the QP weights\cite{Cederbaum1975, Cederbaum1977} and deteriorated performance of many single-reference methods compared to frontier IPs. For instance, for the 23 frontier IPs in the benchmark set of \citet{Marie2024}, CCSD has a MAE of 0.082 eV, while it doubles for second IPs.\cite{Marie2024} Similar observations can be made for (vertex-corrected) $GW$ calculations.\cite{Tolle2026ConnectionCluster} It is therefore interesting to comment on the behaviour of the different ADC approximations in this regard. 

Figure~\ref{fig:z_vs_error_qz} shows the mean absolute percentage and individual percentage errors of the 56 IPs calculated with ADC(3), ADC(2)-X, sMCDE and, for comparison, the $G_0W_0$ approximation (from Ref.~\citenum{Marie2024}) broken down into first, second, and third or higher (3+) IPs. For each category, the average Z-factor is shown, as well as Pearson correlation coefficients. While $G_0W_0$ performs worse for deeper IPs, the performance of ADC(3) remains largely unchanged, whereas ADC(2)-X and sMCDE even systematically improve: the $Z$-factor is not indicative of the accuracy of those methods. Also considering mean absolute errors, ADC(2)-X and sMCDE show nearly flat error distributions for valence, semi-valence and semi-core IPs, even though the latter have larger absolute values.   

\section{Conclusions}
Nowadays, most ADC calculations are performed in the non-Dyson (nD) approximation. This is due to its numerical simplicity as well as due to the fact that past benchmarks have claimed the error introduced by this approximation to be small.\cite{Trofimov2005MolecularApproach} We have here demonstrated at the ADC(3) level that the nD-approximation introduces an error of about 0.1 eV on average for first IPs of 23 small molecules.\cite{Marie2024} We have also shown that Dyson-ADC calculations can be performed readily using non-standard root-following Davidson solvers. We hope that these results will stimulate renewed interest in the Dyson-ADC formalism. 

We have compared the Dyson-ADC and (3,1)-MCDE formalism for the single-particle Green's function. While rooted in different conceptual frameworks, the (3,1)-MCDE working equations are completely equivalent to the ones of the ADC(2)-X method,\cite{Demartini2026AlgebraicEquation} introduced a long time ago by Cederbaum and co-workers\cite{Cederbaum1977, cederbaum1977theoretical, schirmer1978two} as the 2ph-TDA. The (3,1)-MCDE formalism offers a very useful new perspective to incorporate physically motivated techniques into the ADC formalism. Recently, two of the current authors have introduced the sMCDE method\cite{Romaniello2026DirectEquation} that uses the statically screened RPA interaction to screen the ladder terms in the 3-body block of the ADC(2)-X/(3,1)-MCDE Hamiltonian. While this allows the application of the formalism to solids, we have shown here that screening is also beneficial for molecules. This result will not surprise practitioners of the $GW$(-BSE) method that uses screening as a central ingredient.\cite{Onida2002,Golze2019} Out of all methods benchmarked in this paper, the sMCDE has the narrowest error distribution about the mean signed error (the smallest $\sigma^{\textrm{MAE}}$). This should make the sMCDE a useful and cost-efficient tool for the calculation of photoemission spectra in molecules, for which the relative peak positions are often more relevant than absolute errors, and for which promising results have already been obtained with the standard (3,1)-MCDE.\cite{Paggi2026CoreEquation} 

With average errors of about 200-300 meV, Dyson-ADC(3) does not reach the accuracy of vertex-corrected $GW$ approaches or EOM-CCSD. An obvious next avenue of high scientific value would therefore be to investigate how to improve the IPs within Dyson-ADC. A natural approach to try out would be to extend the sMCDE to a screened ADC(3). It is also yet to be established whether the ADC hierarchy converges reliably to FCI. To the best of our knowledge, no variant beyond nD-ADC(4) has so far been implemented and benchmarked and the available results, while indicating robust improvement over nD-ADC(3) especially for satellites,\cite{Banerjee2023AlgebraicSpectra, Leitner2024Fourth-OrderMethods} are not conclusive in this regard. Another interesting question would be whether and how multi-reference (MR) Dyson-ADC can be realized, in analogy to nD-MR-ADC that has been developed by Sokolov and coworkers.\cite{Sokolov2018Multi-referenceImplementation,
Chatterjee2019Second-OrderSystems, Chatterjee2020ExtendedExcitations,
Mazin2021MultireferenceBenchmark,
DeMoura2022SimulatingTheory} Recent work on MR-$GW$ that operates in the Dyson-formalism\cite{Wang2026Multi-referenceMolecules} could offer useful inspiration toward this goal.

% Additional info
\section*{Data and Software Availability Statement}
The data underlying this study are available within the published article and its Supplementary Material. Execution of the AMS input file requires a modified developer version of the code, which is accessible to licensed AMS developers or upon reasonable request. The Python code used to perform the calculations shown in this work is available at \href{https://github.com/ArnoFoerster/MBPTcode}{https://github.com/ArnoFoerster/MBPTcode}

\section*{Supplementary Material}
All data calculated in this work are available as a .csv file. The supplemental information contains a detailed review of the static self-energies and details their implementation for this work.

\section*{Author Declarations} 
The authors declare no conflict of interest.

\section*{Acknowledgments}
The authors acknowledge use of the supercomputer facilities at SURFsara sponsored by Netherlands Organisation for Scientific Research (NWO) Physical Sciences, with financial support from the NWO. AF acknowledges funding through a VENI grant from NWO under grant agreement VI.Veni.232.013.

% References
\bibliography{references.bib,extraRef}
\begin{tocentry}
\includegraphics[width=\textwidth]{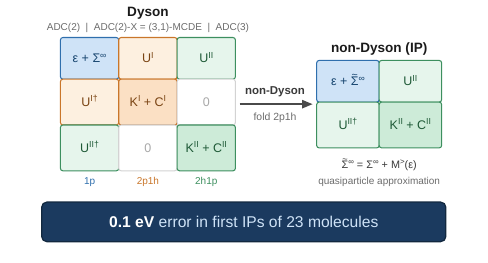}
\end{tocentry}

\end{document}